# Group theory approach to unification of gravity with internal symmetry gauge interactions: I. Canonical electrogravity

S.E. Samokhvalov [a)] and V.S. Vanyashin

a) serg_samokhval@ukr.net

**Abstract.** The infinite group of deformed diffeomorphisms of the spacetime continuum is put into the basis of the gauge theory of gravity. This gives rise to some new ways for unification of gravity with other gauge interactions.

## 1. Introduction

Internal symmetry gauge interactions can be easily unified with each other through the unification of the corresponding symmetry groups. On the contrary, to incorporate gravity into some unified scheme other ways are needed. As an example one could mention supersymmetry, or higher dimensional theories, or both. Such a difference has two reasons. First of all, in the framework of finite dimensional Lie groups a non-trivial unification of the Poincare group with internal symmetry groups is impossible, as the well known Coleman-Mandula no-go theorem states [1]. Secondly, a consistent gauge theory of the gravitation field is still to be found. The widely known gauge approach to gravity, developed in [2,3] in fact gives gauge interpretation neither to metric fields, nor to vierbein ones. An interpretation of these as connection in appropriate fibrings has been achieved at the expense of unnatural juxtapositions [4]. The reason for all these difficulties lies in the fact that the fibre bundles formalism is appropriate only for internal symmetry Lie groups, which do not act on the spacetime manifold, but for gauge gravity this restriction is obviously meaningless.

A generalization of gauge groups for the case of their non-trivial action on the spacetime continuum was proposed in [5]. Gauge fields, corresponding to internal or external (spacetime) symmetry both acquire single interpretation as deformation parameters of infinite dimensional (from now onwards simply infinite) Lie groups.

Giving to the Einstein gravity the consistent status of a gauge theory it allows us to unify gravity with other gauge interactions by the unifications of the initially separated gauge groups into a single one – the deformed infinite Lie group [6]. Such a unification might be a non-trivial one with a good physical content, and for this task much work is certainly needed.

In the present paper the possibilities of the unification of gravity with internal symmetry gauge interactions on the basis of the deformed infinite Lie groups have been demonstrated on the two simple examples of the gravity and electrodynamics unification. The first example, named canonical, leads to the standard Einstein-Maxwell equations in Riemannian spacetime. The second one, in a sense more symmetrical, suggestively leads to some new principles: non-linearity of electrodynamics already on the classical level, inherent smearness of all electric charges, manifestation of the new degree of freedom of the electrogravitational field. The universal Plank length constant $L$ in this case serves for the extension rather than for contraction of physical symmetry. The upper bound on the electrodynamics potentials is also of Plank scale, so for the usually considered non-fantastic field strength values the new example of electrogravity is practically the canonical one.

The first simpler case is presented in the current paper – I. The non-canonical example of electrogravity is considered in II, which will appear later.

## 2. Deformed infinite Lie group

A concise introduction into the approach adopted follows below for convenience. A more detailed description can be found in [5].

Let a Lie group $G$ with the product law $(g_1 \cdot g_2)^\alpha = \tilde{\varphi}^\alpha(g_1, g_2)$ act on the spacetime manifold $M$ (perhaps non-effectively) according to the formula $x'^\mu = \tilde{f}^\mu(x, g)$ [7]. The infinite Lie group $\tilde{\Gamma}_G$ is parametrized by functions $\tilde{g}^\alpha(x)$, which satisfy the condition $\det\{\partial_\nu \tilde{f}^\mu(x, \tilde{g}(x))\} \neq 0$, $\forall x \in M$. The product law in $\tilde{\Gamma}_G$ is determined by the formula:

$$(\tilde{g}_1 \times \tilde{g}_2)^\alpha(x) = \tilde{\varphi}^\alpha(\tilde{g}_1(x), \tilde{g}_2(x')),$$

where $x'^\mu = \tilde{f}^\mu(x, \tilde{g}_1(x))$ sets the action of $\tilde{\Gamma}_G$ on $M$. In the case of trivial action of the group $G$ on $M$, $x'^\mu = x^\mu$ and $\tilde{\Gamma}_G$ appears to be an ordinary gauge group $G^g$.

Let us pass from the group $\tilde{\Gamma}_G$ to the group $\Gamma_G$ isomorphic to it by formula $g^a(x) = H^a(x, \tilde{g}(x))$ (Latin indices assume the same values as the corresponding Greek ones). The functions $H^a(x, g)$ have the properties:

1H  $H^a(x, g) \in C_\infty(M \times G)$

2H  $H^a(x, 0) = 0 \qquad \forall x \in M$

3H  $\exists K^\alpha(x, g): \ K^\alpha(x, H(x, g)) = g^\alpha \quad \forall x \in M, \ g \in G.$

The group $\Gamma_G$ product law is determined by its isomorphism to the group $\tilde{\Gamma}_G$:

$$(g_1 \times g_2)^a(x) := \varphi_x^a(g_1(x), g_2(x')) = H^a(x, \tilde{\varphi}(K(x, g_1(x)), K(x', g_2(x')))), \tag{1}$$

$$x'^\mu = f^\mu(x, g_1(x)) := \tilde{f}^\mu(x, K(x, g_1(x))). \tag{2}$$

Formula (2) sets the action of $\Gamma_G$ on $M$.

Such transformations between the groups $\tilde{\Gamma}_G$ and $\Gamma_G$ we name as deformations of infinite Lie groups, since the corresponding deformations of geometric structures of manifolds, subjected to group action, are directly associated with them [5]. We name the functions $H^a(x, g)$ as deformation functions, and

$$H^a_\alpha(x) := \partial_{g^\alpha} H^a(x, g)\Big|_{g=0}$$

as auxiliary deformation functions or deformation coefficients.

The functions $\varphi_x$, setting the product law (1) in the group $\Gamma_G$, are explicitly $x$ dependent, so the structure coefficients of the group $\Gamma_G$ are $x$ dependent as well:

$$F(x)^a_{bc} := \left(\partial^2_{g_1^b, g_2^c} - \partial^2_{g_1^c, g_2^b}\right)\varphi^a_x(g_1, g_2)\Big|_{g_1=g_2=0}. \tag{3}$$

Since

$$\xi^\mu_a := \partial_{g^a} f^\mu(x, g)\Big|_{g=0} = H^\alpha_a \tilde{\xi}^\mu_\alpha,$$

where $H^\alpha_a$ is reciprocal to the $H^a_\alpha$ matrix, the generators of the group $\Gamma_G$ are expressed by the generators of the group $G$ with the help of deformation coefficients: $X_a := \xi^\mu_a \partial_\mu = H^\alpha_a \tilde{X}_\alpha$, the commutators of which are proportional to the structure functions of the group $\Gamma_G$:

$$[X_a, X_b] = F(x)^c_{ab} X_c. \tag{4}$$

## 3. Gravity as the gauge theory of the translation group

Let $G = T$, where $T$ is the group of spacetime translations. In this case $\tilde{\varphi}^\mu(t_1, t_2) = t^\mu_1 + t^\mu_2$ and $x'^\mu = x^\mu + t^\mu$. The group $\tilde{\Gamma}_T$ is parametrized by the functions $\tilde{t}^\mu(x)$, which satisfy the condition

$$\det\{\delta^\mu_\nu + \partial_\nu \tilde{t}^\mu(x)\} \neq 0 \qquad \forall x \in M.$$

The product law in $\tilde{\Gamma}_T$ is

$$(\tilde{t}_1 \times \tilde{t}_2)^\mu(x) = \tilde{t}^\mu_1(x) + \tilde{t}^\mu_2(x'),$$

where $x'^\mu = x^\mu + \tilde{t}^\mu_1(x)$ determines the action of $\tilde{\Gamma}_T$ on $M$. The product law indicates that $\tilde{\Gamma}_T$ is the group of spacetime diffeomorphisms $\text{Diff} M$ in special parametrization. Thus, in the approach considered, the Diff $M$ group appears to be the gauge group of liberated local translations. The generators of the $\tilde{\Gamma}_T$ action on $M$ are simply derivatives $\tilde{X}_\mu = \partial_\mu$ and this corresponds to the case of the fiat $M$ and absence of gravitational field [5].

Let us deform the group $\tilde{\Gamma}_T \to \Gamma_T$, introducing simultaneously in $M$ gravitational field: $t^m(x) = H^m(x, \tilde{t}(x))$. The product law in $\Gamma_T$ is determined by the formulae:

$$(t_1 \times t_2)^m(x) = H^m(x, K(x, t_1(x)) + K(x', t_2(x'))), \tag{5}$$

$$x'^\mu = x^\mu + K^\mu(x, t_1^\mu(x)). \tag{6}$$

Formula (6) sets the action of $\Gamma_T$ on $M$.

The gravitational field potentials are identified with the auxiliary deformation functions $h_\mu^m(x) = \partial_{t^\mu} H^m(x,t)\big|_{t=0}$, which can be treated as vierbeins. The generators $X_m = h_m^\mu \partial_\mu$ of the $\Gamma_T$ action on $M$ have their commutators proportional to structure functions of the group $\Gamma_T$, which are

$$F_{mn}^k = -h_m^\mu h_n^\nu (\partial_\mu h_\nu^k - \partial_\nu h_\mu^k), \qquad [X_m, X_n] = F_{mn}^k X_k. \tag{7}$$

The structure functions $F_{mn}^k$ differ from non-holonomity coefficients only by a factor $-\frac{1}{2}$ and are identified with the field strength tensor of the gravitational field. The same gravitational field is described by all the groups $\Gamma_T$ obtained one from the other by internal automorphisms. Under these automorphisms, which can always be connected with the transformations of coordinates on $M$, the fields $h_\mu^m$ transform according to the general covariance law, that for the infinitesimal $t^m(x)$ is

$$h'^m_\mu = h_\mu^m + F_{kl}^m h_\mu^l t^k - \partial_\mu t^m \tag{8}$$

and the field strength tensor $F_{mn}^k$ does not change, being a general coordinate scalar.

The transformation law (8) is similar to the transformation law for internal symmetry gauge fields and the only difference consists in the replacement of a finite Lie group structure constants by the infinite Lie group $\Gamma_T$ structure functions. This fact permits us to interpret the group $\Gamma_T$, as the gauge translation group and the vector fields $h_\mu^m$ as the gauge field of the translation group.

The gravitational field action

$$S_h = \tfrac{1}{c} \int L_h(h_\mu^m, \partial_\nu h_\mu^m) \sqrt{-g}\, d^4 x$$

is to be taken $\Gamma_T$–invariant, so that it is to be constructed from invariant field strength tensor $F_{mn}^k$. Among all the bilinear on $F_{mn}^k$ Lagrangians there is one

$$L_h^c = \frac{c^4}{64\pi\gamma}(F_{mn}^k F_{mn}^k + 2F_{mn}^k F_{kn}^m - 4F_{mn}^n F_{mk}^k),$$

which differs from the Einstein Lagrangian only by total divergence [4]. It also acquires a total divergence term under the local Lorentz $\Lambda^g$ transformation, and the $\Lambda^g$ invariance of the gravitational field equations take place. The gravitational field displacement tensor $B_m^{\mu\nu} := \partial_{\partial_\nu h_\mu^m} L_h$ is determined by the choice of $L_h$. $\Gamma_T$ invariance of $L_h$ leads to the following identities of the second Noether theorem:

$$B_m^{\mu\nu} = -B_m^{\nu\mu}, \qquad \sqrt{-g}\, t_m^\mu = -\partial_{h_\mu^m}(\sqrt{-g}\, L_h), \qquad (9)$$

$$\partial_\mu(\sqrt{-g}\, t_m^\mu) = \sqrt{-g}\, \nabla_\mu t_m^\mu = 0, \qquad (10)$$

where $\sqrt{-g}\, t_m^\mu$ is the Noether translation symmetry current, which is conserved on extremals due to weak identity (10). So

$$t_m^\mu = B_n^{\nu\mu} F_{\nu m}^n - L_h h_m^\mu \qquad (11)$$

is the energy-momentum tensor of the gravitational field. The identity (9) is a strong ones. In the presence of matter fields the tensor $t_m^\mu$ in (10) should be replaced by the total energy-momentum tensor $T_m^\mu = t_m^\mu + \tau_m^\mu$, where $\tau_m^\mu$ is the energy-momentum tensor of the gravitating matter. The $\Gamma_T$ invariance, leading to the identities (9) for any choice of $L_h$, gives rise to equations of gravity that are similar to the electromagnetic ones,

$$\nabla_\nu B_n^{\mu\nu} = -t_m^\mu - \tau_m^\mu. \qquad (12)$$

For the special choice $L_h = L_h^c$ equations (12) are the Einstein ones in the vierbein transcription. In this case among the 16 equations (12) there are only six independent ones, because there is the gauge freedom in the choice of $\Gamma_T$ and $\Lambda^g$ gauges, and the 10 gauge conditions should be specified for the gravitational potentials $h_\mu^m$.

## 4. Canonical electrogravity

The form of the gravity equations (12) is uniquely determined by the fact that the gravity is the gauge theory of the translation group. In turn, the electrodynamics is the gauge theory of the phase translation group $U(1)$. So, the suggestive idea to get the electrogravity as the gauge theory of the unified group $\Gamma_{T\otimes U(1)}^c = \Gamma_T \times U(1)^g$ is worth considering.

There are many ways to get a semidirect product of such a kind and all of them are connected by the isomorphisms of the groups $\Gamma_{T \otimes U(1)}$, i.e. by their deformations. Let us consider first the simplest way, which lead to the canonical electrogravity. In this case the deformations of the group $\tilde{\Gamma}_{T \otimes U(1)}$ are chosen so that the auxiliary deformation functions $H_\alpha^a$ have the following triangle form:

$$H_\alpha^a = \begin{pmatrix} h_\mu^m & 0 \\ -lkA_\mu & l \end{pmatrix} \qquad (13)$$

where $l = 2\sqrt{\hbar\gamma/\alpha c^3} = 3.783 \times 10^{-32}$ cm, $k = e/c\hbar$, $\gamma$ is the gravity constant, $\alpha$ is the fine structure constant.

The infinitesimal action on $P = M \times U(1)$ of the corresponding deformed group $\Gamma_{T \otimes U(1)}^c$ is

$$\delta x^\mu = h_m^\mu t^m, \qquad \delta\varphi = -k(c\theta - A_m t^m). \qquad (14)$$

The generators of the transformation (14) are $X_\theta = -kc\partial_\varphi$ and $X_m = h_m^\mu(\partial_\mu + kA_\mu\partial_\varphi)$.

The auxiliary functions $h_\mu^m(x)$ and $A_\mu(x)$ are identified with potentials of gravitational and electromagnetic fields, respectively. The non-trivial structure functions of the group $\Gamma_{T \otimes U(1)}^c$: $F_{mn}^k = -h_m^\mu h_n^\nu(\partial_\mu h_\nu^k - \partial_\nu h_\mu^k)$ and $F_{mn} = h_m^\mu h_n^\nu(\partial_\mu A_\nu - \partial_\nu A_\mu)$ are identified with the field strength tensors of gravitational and electromagnetic fields. The same physical situation is described by all the groups $\Gamma_{T \otimes U(1)}^c$, connected with each other by internal automorphisms, leading to the gauge transformation of the potentials:

$$h_\mu'^m = h_\mu^m + F_{kl}^m h_\mu^l t^k - \partial_\mu t^m, \qquad A_\mu' = A_\mu + F_{\mu\nu} h_k^\nu t^k - c\partial_\mu\theta. \qquad (15)$$

The field strength tensors of gravitational and electromagnetic fields are gauge invariant, of course, and may produce a $\Gamma_{T \otimes U(1)}^c$ invariant Lagrangian of canonical electrogravity:

$$L_{hA}^c = L_h^c - \frac{1}{16\pi} F^{kl} F_{kl}. \qquad (16)$$

The $\Gamma_{T \otimes U(1)}^c$ invariance determines the following form of the equations of the canonical electrogravity:

$$\nabla_\nu F^{\mu\nu} = -\frac{4\pi}{c} j^\mu, \qquad \nabla_\nu B_n^{\mu\nu} = -t_m^\mu - \theta_m^\mu - \tau_m^\mu, \qquad (17)$$

where

$$\theta_m^\mu = F^{\nu\mu} F_{\nu m} - \frac{1}{4} h_m^\mu F^{\nu\rho} F_{\nu\rho}$$

is the energy-momentum tensor of an electromagnetic field.

## Acknowledgments

One of the authors (VSV) would like to thank Professors Abdus Salam, Luciano Bertocch, Kumar Narain and James Eells for their kind hospitality at the International Centre for Theoretical Physics, Trieste, where final part of this work was done.